\documentclass[12pt,preprint]{aastex}
\def\degpoint{\ifmmode ^{\rm{o}}\!. \else $^{\rm{o}}\!.$\fi}

\newcommand{\ms}{\mbox{m\,s$^{-1}$}}
\newcommand{\kms}{\mbox{km \ s$^{-1}$}}
\newcommand{\Msun}{\mbox{M$_{\odot}$}}
\newcommand{\Rsun}{\mbox{R$_{\odot}$}}
\newcommand{\Mjup}{\mbox{M$_{\rm Jup}$}}

\newcommand{\Lsun}{\mbox{L$_{\odot}$}}

\newcommand{\gtsimeq}{\raisebox{-0.6ex}{$\,\stackrel
         {\raisebox{-.2ex}{$\textstyle >$}}{\sim}\,$}}

\begin{document}

\title{The Pan-Pacific Planet Search. IV. Two super-Jupiters in a 3:5 
resonance orbiting the giant star HD\,33844 }

\author{Robert A.~Wittenmyer\altaffilmark{1,2,5}, John Asher 
Johnson\altaffilmark{3}, R.P. Butler\altaffilmark{4}, Jonathan 
Horner\altaffilmark{2,5}, Liang Wang\altaffilmark{6}, Paul 
Robertson\altaffilmark{7,8,9}, M.I.~Jones\altaffilmark{10}, 
J.S.~Jenkins\altaffilmark{11}, R. Brahm\altaffilmark{12}, 
C.G.~Tinney\altaffilmark{1,2}, M.W. Mengel\altaffilmark{5}, 
J.~Clark\altaffilmark{13} }

\altaffiltext{1}{School of Physics, University of New South Wales, 
Sydney 2052, Australia}
\altaffiltext{2}{Australian Centre for Astrobiology, University of New 
South Wales, Sydney 2052, Australia}
\altaffiltext{3}{Harvard-Smithsonian Center for Astrophysics, Cambridge,
MA 02138 USA}
\altaffiltext{4}{Department of Terrestrial Magnetism, Carnegie 
Institution of Washington, 5241 Broad Branch Road, NW, Washington, DC 
20015-1305, USA}
\altaffiltext{5}{Computational Engineering and Science Research Centre, 
University of Southern Queensland, Toowoomba, Queensland 4350, 
Australia}
\altaffiltext{6}{Key Laboratory of Optical Astronomy, National
Astronomical Observatories, Chinese Academy of Sciences, A20 Datun Road,
Chaoyang District, Beijing 100012, China}
\altaffiltext{7}{NASA Sagan Fellow}
\altaffiltext{8}{Department of Astronomy and Astrophysics, The 
Pennsylvania State University, USA}
\altaffiltext{9}{Center for Exoplanets \& Habitable Worlds, The 
Pennsylvania State University, USA}
\altaffiltext{10}{Department of Electrical Engineering and Center of 
Astro-Engineering UC, Pontificia Universidad Cato\'lica de Chile, Av. 
Vicuña Mackenna 4860, 782-0436 Macul, Santiago, Chile}
\altaffiltext{11}{Departamento de Astronomi\'a, Universidad de Chile, 
Camino El Observatorio 1515, Las Condes, Santiago, Chile}
\altaffiltext{12}{Instituto de Astrofi\'sica, Facultad de Fi\'sica, 
Pontificia Universidad Cato\'lica de Chile and Millennium Institute of 
Astrophysics, Av. Vicun\~a Mackenna 4860, 7820436 Macul, Santiago, 
Chile}
\altaffiltext{13}{School of Physical Sciences, University of Adelaide, 
Adelaide SA 5005, Australia}

\email{
rob@unsw.edu.au}

\shorttitle{HD\,33844 two-planet system }
\shortauthors{Wittenmyer et al.}

%-------------------------------------------------------------------
\begin{abstract}

\noindent We report the discovery of two giant planets orbiting the K 
giant HD\,33844 based on radial velocity data from three independent 
campaigns.  The planets move on nearly circular orbits with semimajor 
axes $a_b=1.60\pm$0.02\,AU and $a_c=2.24\pm$0.05\,AU, and have minimum 
masses (m sin $i$) of $M_b=1.96\pm$0.12\,\Mjup\ and 
$M_c=1.76\pm$0.18\,\Mjup.  Detailed N-body dynamical simulations show 
that the two planets remain on stable orbits for more than $10^6$ years 
for low eccentricities, and are most likely trapped in a mutual 3:5 
mean-motion resonance.

\end{abstract}


%--------------------------------------------------------------------
\section{Introduction}

Surveys for planets orbiting evolved stars more massive than the Sun are 
well into their second decade.  The longest-running surveys 
\citep[e.g.][]{sato05, reffert15} have been monitoring several hundred 
such stars for $\sim$15 years.  The combined efforts of these and other 
surveys have amassed enough data to begin making quantitative statements 
about the frequency and detailed properties of planetary systems beyond 
solar-type main-sequence stars.

An early prediction from formation models proposed that higher-mass 
stars should host higher-mass planets \citep{idalin05}, a prediction 
that is being borne out by observation \citep{bowler10}.  Giant planet 
frequency has also been shown to increase with host-star mass 
\citep{fv05, johnson10, bowler10}, though with a drop-off for hosts with 
$M_{*}>2.5-3.0$\,\Msun\ \citep{omiya09, kunitomo11, reffert15}.  
\citet{kretke09} proposed a mechanism to explain the efficient formation 
of gas giant planets at orbital distances $a\gtsimeq$1\,AU.  For 
intermediate-mass stars, the inner edge of the magneto-rotational 
instability (MRI) dead zone lies far enough from the star to permit 
cores to accrete gas rapidly, producing gas giants at a higher rate than 
for solar-mass stars.  An interesting consequence of their models is 
that the frequency of giant planets would have little dependence on 
stellar metallicity, in contrast to the well-known planet-metallicity 
correlation for dwarf stars \citep{fv05}.  However, recent results from 
\citet{reffert15}, with a sufficiently large and self-consistent sample 
of intermediate-mass stars and their planets in hand, show that planet 
occurrence remains positively correlated with metallicity for these 
stars.

The Pan-Pacific Planet Search (PPPS - Wittenmyer et al.~2011b) was a 
radial velocity survey of 170 Southern giant stars using the 3.9m 
Anglo-Australian Telescope (AAT) and its UCLES high-resolution 
spectrograph \citep{diego:90}.  It was originally conceived as a 
Southern hemisphere extension of the Lick \& Keck Observatory survey for 
planets orbiting Northern ``retired A stars'' \citep{johnson06b}.  The 
targets were selected to be redder ($1.0 < (B-V) < 1.2$) than the 
Northern hemisphere sample in order to select for more metal-rich stars 
\citep{girardi02}.  The PPPS operated from 2009-2014; papers detailing 
the spectroscopic stellar parameters and new planet detections are now 
in preparation \citep{155233, stellar}.  This paper is organised as 
follows: Section 2 details the AAT and Keck observations of HD\,33844 
and gives the stellar parameters.  Section 3 describes the orbit-fitting 
procedures and gives the parameters of the two planets in the HD\,33844 
system.  In Section 4 we discuss the evidence for a planetary 
interpretation of the observed radial velocity variations, including 
dynamical stability simulations.  Then we give our conclusions in 
Section 5.

%--------------------------------------------------------------------
\section{Observations and Stellar Properties}

HD\,33844 is common to the AAT, Keck, and FEROS evolved-star surveys.  
Precision Doppler measurements for the PPPS are obtained with the UCLES 
echelle spectrograph at the AAT.  The observing procedure is identical 
to that used by the long-running Anglo-Australian Planet Search 
\citep[e.g.][]{tinney01,butler01,jones10,142paper}; a 1-arcsecond slit 
delivers a resolving power of $R\sim$45,000.  Calibration of the 
spectrograph point-spread function is achieved using an iodine 
absorption cell temperature-controlled at 60.0$\pm$0.1$^{\rm{o}}$C.  The 
iodine cell superimposes a forest of narrow absorption lines from 5000 
to 6200\,\AA, allowing simultaneous calibration of instrumental drifts 
as well as a precise wavelength reference \citep{val:95,BuMaWi96}.

We have obtained 20 AAT observations of HD\,33844 since 2009 Feb 4, and 
an iodine-free template spectrum was obtained on 2011 January 19.  With 
$V=7.29$, exposure times are typically 900-1200\,s, with a resulting S/N 
of $\sim$100-200 per pixel each epoch.  The data, given in 
Table~\ref{aatvels}, span a total of 1880 days (5.5\,yr), and have a 
mean internal velocity uncertainty of 2.1\,\ms.  

HD\,33844 was also observed with the High Resolution Echelle 
Spectrometer (HIRES) on the 10m Keck I telescope.  A total of 36 epochs 
have been obtained, spanning 2190 days (6 yr).  Radial velocities were 
computed using the iodine-cell method as described above; the data are 
given in Table~\ref{keckvels} and have a mean internal uncertainty of 
1.3\,\ms.

%A 0.86-arcsecond slit
%delivers a resolution of $R\sim\,55,000$.
%An exposure meter ensures 
%that each observation \textbf{has S/N of YYY per pixel in the iodine 
%region (5800\AA).  Typical exposure times ranged from XXX-YYY seconds.}  

We also include 11 radial velocity observations from the FEROS 
spectrograph \citep{kaufer99} on the 2.2m telescope at La Silla 
Observatory.  Those data are part of the EXPRESS (EXoPlanets aRound 
Evolved StarS) survey \citep{jones11, jones15} for planets orbiting 
evolved stars.  The PPPS and EXPRESS surveys have 37 targets in common; 
further papers are in preparation detailing joint planet discoveries 
made possible by the combination of the two data sets.  The FEROS data 
for HD\,33844 are given in Table~\ref{ferosvels}; they cover a span of 
1108 days and have a mean internal uncertainty of 3.9\,\ms.  The typical 
observing time was 250\,s, leading to a S/N of 200 per pixel.  The 
spectra were reduced using a flexible pipeline for echelle spectra 
(Jordan et al. 2014; Brahm et al. 2015, in preparation). The radial 
velocities were computed using the simultaneous calibration technique, 
according to the method described in \citet{jones13} and \citet{jj14}.

%spectra were reduced using the FEROS piepline, and the radial velocities 
%were computed using the simultaneous calibration method 
%\citep{baranne96}, according to the method described in \citet{jones13} 
%and \citet{jj14}.

\subsection{Stellar Properties}

%  this is stolen from earlier papers

We have used our iodine-free template spectrum ($R\sim$60,000, 
S/N$\sim$200) to derive spectroscopic stellar parameters.  In brief, the 
iron abundance [Fe/H] was determined from the equivalent widths of 32 
unblended Fe lines, and the LTE model atmospheres adopted in this work 
were interpolated from the ODFNEW grid of ATLAS9 \citep{Castelli2004}.  
The effective temperature ($T_\mathrm{eff}$) and bolometric correction 
($BC$) were derived from the color index $B-V$ and the estimated 
metallicity using the empirical calibration of 
\citet{Alonso1999,Alonso2001}.  Since the color-$T_\mathrm{eff}$ method 
is not extinction-free, we corrected for reddening using $E(B-V)=0.0290$ 
\citep{schlegel98}.  The stellar mass and age were estimated from the 
interpolation of Yonsei-Yale ($\mathrm{Y}^2$) stellar evolution tracks 
\citep{Yi2003}.  The resulting stellar mass of 1.78$\pm$0.18\,\Msun\ was 
adopted for calculating the planet masses.  Our derived stellar 
parameters are given in Table~\ref{stellarparams}, and are in excellent 
agreement with the results of \citep{jones11}, who found a mass of 
1.74$\pm$0.18\,\Msun\ and radius 5.33$\pm$0.51\,\Rsun.

%--------------------------------------------------------------------
\section{Orbit Fitting and Planetary Parameters}

Early AAT data for HD\,33844 exhibited a periodicity of $\sim$510 days, 
but the one-planet fit worsened with time until it could be tentatively 
fit with a second planet near $\sim$900 days.  Preliminary analysis of 
the AAT and Keck data together corroborated the two candidate 
periodicities.  We first explored a wide range of parameter space by 
fitting the two data sets with a two-Keplerian model within a genetic 
algorithm \citep[e.g.][]{HUAqr, NNSer, songhu}.  In brief, the 
genetic algorithm works on principles of evolutionary biology, producing 
an initially random population of planetary system parameters, then 
selecting the best-fit (lowest $\chi^2$) models for ``reproduction.'' 
The next generation is then generated by perturbing the best-fit models 
(``mutation'') and repeating the process.  The two planets were allowed 
to take on orbital periods in the range $P_{1}: 400-600$d and $P_{2}: 
700-1200$d, and eccentricities $e<0.3$.  A total of about $10^7$ 
possible system configurations were tested in this manner.  The best 
two-planet solution was then used as a starting point for the 
generalized least-squares program \textit{GaussFit} \citep{jefferys88}, 
here used to solve a Keplerian radial-velocity orbit model as in our 
previous work \citep{tinney11,47205paper,121056}.  As a further check, 
we performed a Keplerian fit, optimised with a simplex 
algorithm, using version 2.1730 of the \textit{Systemic Console}  
\citep{mes09} and estimated parameter uncertainties using the bootstrap 
routine therein on 100,000 synthetic data set realisations.  We 
added 7\,\ms\ of jitter in quadrature to the uncertainties of each of 
the three data sets.  This jitter estimate is derived from 37 stable 
stars in the PPPS (334 measurements); their velocity distribution can be 
fit with a Gaussian of width $\sigma\,=7$\,\ms.  Since the planets are 
massive and move on orbits relatively close to each other (such that 
interations can be expected), we also performed a dynamical fit using 
the Runge-Kutta integration method within \textit{Systemic}.  
Table~\ref{planetparams} gives the planetary system parameters resulting 
from both the Keplerian and dynamical fits; the results are 
indistinguishable and hence neither technique is clearly favoured.  The 
parameters given represent the mean of the posterior distribution and 
the 68.7\% confidence interval.  Using the host star mass of 
1.78\,\Msun\ in Table~\ref{stellarparams}, we derive planetary minimum 
masses of 1.96$\pm$0.12\,\Mjup\ (HD\,33844b) and 1.76$\pm$0.18\,\Mjup\ 
(HD\,33844c).  The data and model fits for each planet are plotted in 
Figures 1-2.

%We also plot the data phased to the orbital period of each 
%planet -- the absence of large phase gaps adds confidence to the 
%trustworthiness of the two-planet model fits.

%--------------------------------------------------------------------
\section{Discussion}

\subsection{Evidence for orbiting planets}

Particularly for giant stars, where spots and pulsations can induce 
spurious radial velocity shifts with periods of hundreds of days, any 
claim of orbiting planets must be carefully examined to rule out 
intrinsic stellar signals \citep[e.g.][]{hatzes00, reffert15, 
trifonov15}.  For HD\,33844, the periods of the two signals (551 and 916 
days) are nowhere near the window function peaks at 384 and 8.1 days 
(AAT) or 30 and 364 days (Keck).  Spurious periods in observational data 
commonly arise at those periods due to sampling (imposed by bright-time 
scheduling and the yearly observability of a given target).

To check whether the observed velocity variations could be due to 
intrinsic stellar processes, we examined the All-Sky Automated Survey 
(ASAS) $V$ band photometric data for HD\,33844 \citep{asas}.  A total of 
596 epochs were obtained from the ASAS All Star 
Catalogue\footnote{http://www.astrouw.edu.pl/asas}.  We computed the 
mean magnitude per epoch over the five apertures, then subjected the 
time series to an iterative sigma-clipping process.  We removed points 
more than $3\sigma$ from the grand mean, then recalculated the mean and 
its standard deviation.  This process was performed three times, after 
which 511 epochs remained with a mean value of 7.28$\pm$0.02.  The 
generalised Lomb-Scargle periodogram \citep{zk09} is shown in 
Figure~\ref{asas_pgram}, with the periods of the planets marked as 
dashed lines.  While there are significant periodicities at 730 and 1250 
days, there is little power near the periods of the candidate planets 
(551 and 916 days).

We also checked for correlation between the radial velocities and the 
equivalent width of the H$\alpha$ absorption line, which has been used 
as an activity indicator for giants \citep{hatzes15} as well as for M 
dwarfs \citep{robertson13}.  The equivalent widths were measured in a 
2\,\AA\ window centered on H$\alpha$ to avoid contamination by telluric 
lines.  A generalized Lomb-Scargle periodogram of the H$\alpha$ 
equivalent widths from the AAT spectra (Figure~\ref{halpha}) shows no 
significant periodicities, and there are no correlations with the 
velocities.  Furthermore, the bisector velocity spans (defined as the 
velocity difference between line bisectors from the upper and lower part 
of an absorption line) computed from the FEROS spectra show no 
correlation with the radial velocities.

\subsection{Dynamical stability}

The HD\,33844 system appears to contain two super-Jupiter planets in 
orbits relatively close to each other.  Given their mass and 
proximity, it is clearly important to consider whether the planets are 
dynamically feasible.  Tha is, could planets on such tightly packed 
orbits be dynamically stable on timescales comparable to the lifetime of 
the system?  A first estimate of the system's stability can be garnered 
by simply assessing the dynamical separation of the two planets, 
considering the separation of their orbits compared to their mutual Hill 
radius.  Following \citet{gladman93}, we can calculate the mutual Hill 
radius of the two planets as follows: 

\begin{equation}
R_H = \Big{[}\frac{(m_1 + m_2)}{3 \Msun}\Big{]}^{1/3} \Big{[}\frac{(a_1 
+ a_2)}{2}\Big{]} ,
\end{equation}

\noindent where the symbols have their usual meaning, and the 
subscripts refer to the inner (1) and outer (2) planets respectively.  
Following this formulism, we find that the best-fit orbits for the two 
candidate planets are separated by 3.8 times their mutual Hill radius 
($R_H = 0.167$\,AU).  For low-eccentricity orbits, \citet{gladman93} 
found that orbits beame unstable at separations smaller than 
$\sim\,2\sqrt{3} = 3.46\,R_H$.  The HD 33844 system is therefore close 
to this critical separation, and as the proposed orbits are somewhat 
eccentric, it is clearly important to subject to the proposed planets to 
further scrutiny.  In contrast to widely-separated systems such as 
HD\,121056, (where the planets orbit at 0.4 and 3.0\,AU - more than 9 
mutual Hill radii apart), for which N-body simulations were not 
necessary, here we must rigorously test the HD\,33844 system stability 
as in our previous work \citep[e.g.][]{marshall10, NSVS, QSVir}.

Most interesting are those systems \citep[e.g.][]{texas1, texas2, 
subgiants} for which the planets prove stable and dynamically feasible 
across just a small fraction of the potential orbital solutions.  In 
these cases, which typically feature planets moving on, or close to, 
mutually resonant orbits, dynamical simulations serve a dual purpose.  
First, they provide evidence that supports the existence of the planets, 
and second, they provide a strong additional constraint on the potential 
orbits followed by those planets, helping us to better tie down their 
true orbits than can be achieved on the basis of the observations alone.

Here, we study the stability of the candidate planets orbiting HD\,33844 
following a now well-established route.  We created a suite of 126,075 
copies of the HD\,33844 system.  In each of these cloned systems, the 
initial orbit of HD\,33844b was the same, located at its nominal 
best-fit values (Table~\ref{planetparams}).  For each system, we 
systematically varied the initial semi-major axis ($a$), eccentricity 
($e$), argument of periastron ($\omega$), and mean anomaly (M) of 
HD\,33844c.  The masses of the planets were held fixed at their 
minimum values (m~sin~$i$, Table~\ref{planetparams}).  We note that 
changing the mass of the planets could alter the stability of the 
system.  This can be illustrated by examination of Equation 1 - it is 
immediately apparent that if the masses of the planets are increased, so 
too is the size of their mutual Hill radius, and thereby the strength of 
their mutual interaction.  However, the effect is actually relatively 
small when compared to the influence of their orbital elements.  As 
such, in this work, we solely explore the influence of `element space', 
and leave the exploration of `mass-space' for future work, once the 
orbital elements of the planets have been better constrained through 
follow-up observations.  Once the uncertainties in those elements are 
sufficiently small, it might be possible to use `mass-space' to 
constrain the maximum masses of the planets, and thereby obtain some 
constraints on the inclination of the system to our line of sight, but 
such calculations are beyond the scope of this work.  We do, however, 
note that a tentative upper limit on the masses of the planets can be 
obtained using the resonance overlap criterion \citep[e.g.][]{wisdom80, 
deck13}.  This analytic estimate sets an upper bound of $\sim$10\,\Mjup 
for each planet.  As such, we can be fairly confident that the two 
bodies are planetary in nature, rather than being brown dwarfs.

Since previous studies have shown that the stability of a system is most 
strongly dependent on semi-major axis and eccentricity, we tested 41 
discrete values of each of these variables, spanning the full 
$\pm\,3\sigma$ uncertainty ranges.  At each of the 1681 $a-e$ pairs 
created in this way, we tested fifteen unique values of the argument of 
periastron, and five of the mean anomaly, distributed in each case 
evenly across the $1\sigma$ uncertainty ranges in these variables.  In 
total, then, this gave us 126,075 unique potential orbits for 
HD\,33844c.

We then used the Hybrid integrator within the n-body dynamics package 
\textsc{Mercury} \citep{chambers99} to follow the evolution of each of 
the test planetary systems for a period of 100 Myrs.  Simulations were 
stopped early if either of the planets were ejected from the system 
(upon reaching a barycentric distance of 5\,AU, which would require 
significant strong instability between the two planets).  They were also 
halted if either of the planets fell into the central star, whose mass 
was set at 1.78\,\Msun\, or if they collided with one another.  If any 
of these events happened, the time of collision/ejection was recorded, 
and the simulation was brought to a close.

As a result of these simulations, we are able to examine the dynamical 
stability of the HD\,33844 system as a function of the initial orbit on 
which HD\,33844c was placed.  Figure~\ref{dynamics1} shows the stability 
of the system as a function of the initial semi-major axis and 
eccentricity of that planet's orbit.  In Figure~\ref{chi}, we show how 
the orbital solutions tested in our dynamical simulations fit to the 
observed data, expressed in terms of the difference in total $\chi^2$ 
relative to the best fit.

%Nearly all of the parameter space tested falls within 
%$\chi^2_{\nu}\,<4$.

It is immediately clear from Figure~\ref{dynamics1} that the proposed 
orbital solution for the system lies in a region of complex dynamical 
behavior, with both extremely stable and unstable solutions being 
possible.  It is reassuring, however, to note that broad regions of 
dynamical stability lie comfortably within the 1-sigma uncertainties on 
the proposed solution - particularly towards lower eccentricities.  We 
note that least-squares radial-velocity fitting routines are well-known 
to inflate eccentricities \citep[e.g.][]{shen08, otoole09a, songhu}.  
That stability is due to orbits in that region being trapped in mutual 
3:5 mean motion resonance with the orbit of HD\,33844b.  From 
Table~\ref{planetparams}, the Keplerian solution gives a period ratio of 
1.661, and 1.687 for the dynamical fit.  Within their uncertainties, 
these solutions agree with each other and are wholly consistent with the 
3:5 resonance (period ratio 1.667).

In addition to the central region of stability, orbits at smaller 
semi-major axes fall into a broad region of stability that extends 
across the full span of tested orbital eccentricities. This feature is 
the result of the mutual 2:3 mean-motion resonance between HD\,33844c 
and HD\,33844b, which is centered on 2.102\,AU.

We can also see evidence of unstable resonant behavior through the plot.  
Most strikingly, there is a band of unstable solutions centered at 
2.33\,AU.  This band is the result of the 4:7 mean motion resonance 
between the two planets.  A further unstable resonant region can be seen 
around 2.195\,AU, associated with the 5:8 resonance between the planets. 
Finally, the 5:9 resonance can be found at 2.375\,AU, which is likely 
the cause of the sculpting of the stability of the system in that 
region.

As such, we can conclude that the candidate planets orbiting HD\,33844 
are dynamically feasible, but that they most likely move on mutually 
resonant, low-eccentricity orbits.  As a further check, we 
investigated the behaviour of the resonant angles for a number of key 
resonances, in the vicinity of the complicated `stability terrain' 
around the best fit orbit.  We found that the best fit orbital solution 
is strongly influenced by its proximity to the 3:5 resonance.  In 
particular, we found that the resonant angle $\phi = 5\lambda_2 - 
3\lambda_1 - \omega_1 + \omega_2$ alternates between libration and 
circulation in a regular manner, completing three full cycles (one 
libration + one circulation) per 500 years of integration.  A similar, 
but noisier, behaviour was observed for a scenario in which the planets 
were located on orbits whose periods were in 5:8 commensurability.  
Here, the resonant angle for the 5:8 mean motion resonance switched 
chaotically between periods of libration (lasting up to a thousand 
years) and period of smooth, slow circulation.  The influence of 
resonant interaction for these solutions was unmistakable, and given the 
proximity of the best-fit solution to the location of the 3:5 resonance, 
it seems most likely that the planets are trapped within it (although we 
note that the 5:8 and 7:12 mean-motion resonances also fall within the 
1$\sigma$ uncertainty in semi-major axis), together with an abundance of 
higher-order, weaker resonances.

%--------------------------------------------------------------------
\section{Conclusions}

We have given evidence for two super-Jovian mass planets orbiting the 
metal-rich ([Fe/H]$= +0.27\pm$0.09) giant HD\,33844.  This result is 
consistent with findings from \citet{reffert15} and \citet{maldonado13} 
demonstrating that metal-rich stars with masses greater than 1.5\,\Msun\ 
are more likely to host planets.  To date, relatively few systems of 
multiple giant planets are known to orbit evolved stars.  
Figure~\ref{multis} shows the 12 previously known multiple-planet 
systems orbiting evolved stars (log $g<4.0$).  The HD\,33844 system is 
included as large red circles.  HD\,33844 is also a multiple-Jovian 
planet system in which all of the gas giants (m sin $i>0.2$\,\Mjup) have 
low eccentricities ($e<0.2$).  Such a configuration is relatively 
uncommon \citep{harakawa15}, with only 15 systems known to date.  
\citet{jones15} noted that of the multiple-planet systems known to orbit 
evolved stars, all but one of the host stars were first-ascent giants; 
HD\,33844 adds to this count as it is near the base of the red giant 
branch.  This is relevant since, although the inner planet has a large 
orbital distance ($a\sim\,1.6$ AU), it might eventually be engulfed in a 
distant future due to tidal interaction with the host star, while the 
outer planet ($a\sim\,2.3$ AU) might eventually survive such a process 
\citep[e.g.][]{v07, kunitomo11, mustill12}.  As a result, in a distant 
future, this system might evolve to a single-planet system, which is 
what we typically find around post-RGB stars, as suggested by 
\citet{jones15}.

It has been noted by \citet{ghezzi10} and \citet{sousa08} that there may 
be a correlation between the stellar metallicity and the masses of the 
planets, i.e. stars hosting only $\sim$\,Neptune-mass planets tend to 
have lower metallicity than stars hosting Jupiter-mass planets.  In 
particular, \citet{ghezzi10} remarked that it is possible that 
``metallicity plays an important role in setting the mass of the most 
massive planet.'' We have checked for the possibility of additional 
undetected planets using our well-established detection-limit methods 
\citep[e.g.][]{limitspaper,jupiters,pasp}.  For our data on HD\,33844, 
with a total residual rms of 7.3\,\ms, we can rule out the presence of 
additional planets with m~sin~$i>0.3$\,\Mjup\ interior to HD\,33844b at 
99\% confidence.  To push this limit down to the Neptune-mass regime, 
one must observe at higher cadence \citep{weihai} or adopt observing 
strategies specifically intended to mitigate stellar oscillation noise 
\citep{otoole08, dumusque11}.

% plot planet masses vs stellar metallicity
%  shows nothing of interest
%planet mass vs star mass
% evolved stars only (log g < 4.0)
% still shows nothing of interest

%--------------------------------------------------------------------
\acknowledgements

JH is supported by USQ's Strategic Research Fund: the STARWINDS project.  
CGT is supported by Australian Research Council grants DP0774000 and 
DP130102695.  We gratefully acknowledge the efforts of PPPS guest 
observers Brad Carter, Hugh Jones, and Simon O'Toole.  This research has 
made use of NASA's Astrophysics Data System (ADS), and the SIMBAD 
database, operated at CDS, Strasbourg, France.  This research has also 
made use of the Exoplanet Orbit Database and the Exoplanet Data Explorer 
at exoplanets.org \citep{wright11}.

%--------------------------------------------------------------------

%----------------------------------------------------------
\begin{deluxetable}{lrr}
\tabletypesize{\scriptsize}
\tablecolumns{3}
\tablewidth{0pt}
\tablecaption{AAT Radial Velocities for HD 33844}
\label{aatvels}
\tablehead{
\colhead{BJD-2400000} & \colhead{Velocity (\ms)} & \colhead{Uncertainty
(\ms)}}
\startdata
54867.00962  &    -14.89  &    1.63  \\
55139.20744  &    -35.49  &    2.00  \\
55525.14921  &     -2.27  &    1.97  \\
55580.04370  &    -36.05  &    1.83  \\
55601.94622  &    -51.59  &    1.79  \\
55879.18697  &      4.37  &    2.28  \\
55880.14794  &      3.28  &    1.75  \\
55881.12234  &     -1.95  &    1.93  \\
55906.00994  &     13.02  &    2.06  \\
55968.97219  &     13.41  &    1.53  \\
55993.93352  &      8.92  &    3.17  \\
56051.85691  &      8.61  &    3.18  \\
56343.96912  &     -1.05  &    2.30  \\
56374.91336  &      3.36  &    1.82  \\
56376.92296  &      3.92  &    1.79  \\
56377.91375  &      0.00  &    1.44  \\
56399.92598  &     16.56  &    2.37  \\
56530.28448  &      8.68  &    2.28  \\
56685.94081  &    -57.81  &    2.34  \\
56747.87554  &    -64.48  &    2.16  \\
\enddata
\end{deluxetable}

%----------------------------------------------------------

\begin{deluxetable}{lrr}
\tabletypesize{\scriptsize}
\tablecolumns{3}
\tablewidth{0pt}
\tablecaption{Keck Radial Velocities for HD 33844}
\label{keckvels}
\tablehead{
\colhead{BJD-2400000} & \colhead{Velocity (\ms)} & \colhead{Uncertainty
(\ms)}}
\startdata
54340.13214  &      29.1  &    1.2  \\
54400.03609  &      20.0  &    1.2  \\
54461.88282  &       3.4  &    1.4  \\
54718.14825  &     -19.7  &    1.3  \\
54791.07499  &     -10.8  &    1.4  \\
54809.92924  &      -1.6  &    1.3  \\
54839.01997  &      -0.4  &    1.4  \\
54846.97020  &      -0.7  &    1.5  \\
54864.91819  &       4.6  &    1.4  \\
54929.72286  &     -11.5  &    1.5  \\
55079.13055  &     -43.3  &    1.3  \\
55109.10676  &     -31.3  &    1.3  \\
55173.05357  &     -22.4  &    1.2  \\
55187.90328  &     -13.5  &    1.3  \\
55197.97134  &     -12.3  &    1.4  \\
55229.77649  &       0.2  &    1.3  \\
55255.74938  &      20.3  &    1.3  \\
55285.78110  &      47.9  &    1.3  \\
55312.72317  &      52.1  &    1.4  \\
55428.13474  &       9.6  &    1.2  \\
55456.04501  &      22.6  &    1.2  \\
55490.96109  &      12.7  &    1.4  \\
55521.97151  &      -3.1  &    1.4  \\
55546.07504  &     -13.0  &    1.4  \\
55584.91662  &     -35.2  &    1.3  \\
55633.81481  &     -58.0  &    1.4  \\
55791.13774  &     -33.8  &    1.2  \\
55810.13994  &     -19.4  &    1.1  \\
55902.01080  &       9.5  &    1.2  \\
55904.86470  &      10.1  &    1.4  \\
55931.98977  &       9.7  &    1.3  \\
55960.77092  &      26.9  &    1.3  \\
55972.77806  &      10.9  &    1.3  \\
56197.06671  &     -18.9  &    1.3  \\
56319.74545  &       7.3  &    1.4  \\
56530.11001  &      14.2  &    1.3  \\
\enddata
\end{deluxetable}

%----------------------------------------------------------

\begin{deluxetable}{lrr}
\tabletypesize{\scriptsize}
\tablecolumns{3}
\tablewidth{0pt}
\tablecaption{FEROS Radial Velocities for HD 33844}
\label{ferosvels}
\tablehead{
\colhead{BJD-2400000} & \colhead{Velocity (\ms)} & \colhead{Uncertainty
(\ms)}}
\startdata
55457.83090  &      39.0  &    5.2  \\
55612.57290  &     -55.3  &    3.8  \\
56160.93800  &      -1.4  &    4.0  \\
56230.78520  &     -10.3  &    3.9  \\
56241.78170  &      -7.2  &    4.5  \\
56251.83720  &     -23.6  &    2.6  \\
56321.60110  &      15.8  &    3.7  \\
56331.62140  &      24.1  &    3.9  \\
56342.58360  &      23.9  &    3.5  \\
56412.47550  &       8.8  &    4.6  \\
56565.79190  &     -13.7  &    3.3  \\
\enddata
\end{deluxetable}

%----------------------------------------------------------

\begin{deluxetable}{lll}
\tabletypesize{\scriptsize}
\tablecolumns{3}
\tablewidth{0pt}
\tablecaption{Stellar Parameters for HD 33844}
\tablehead{
\colhead{Parameter} & \colhead{Value} & \colhead{Reference}
 }
\startdata
\label{stellarparams}
Spec.~Type & K0 III & \citet{houk88} \\
Distance (pc) & 100.9$\pm$6.5 & \citet{vl07} \\
$(B-V)$ & 1.040$\pm$0.009 & \citet{perryman97} \\
$E(B-V)$ & 0.0290 & \\
$A_V$ & 0.0903 & \\
Mass (\Msun) & 1.78$\pm$0.18 & This work \\
   & 1.74$\pm$0.18 & \citet{jones11} \\
V sin $i$ (\kms) & $<$1 & This work \\
   & 1.65 & \citet{jones11} \\
$[Fe/H]$ & +0.27$\pm$0.09 & This work \\
   & +0.17$\pm$0.10 & \citet{jones11} \\
   & +0.19$\pm$0.12 & \citet{luck07} \\
$T_{eff}$ (K) & 4861$\pm$100 & This work \\
   & 4890 & \citet{jones11} \\
   & 4710 & \citet{mass08} \\
   & 4886 & \citet{luck07} \\
log $g$ & 3.24$\pm$0.08 & This work \\
   & 3.05 & \citet{jones11} \\
   & 3.1 & \citet{mass08} \\
$v_t$ (\kms) & 1.00$\pm$0.15 & This work \\
   & 1.17 & \citet{jones11} \\
   & 1.42 & \citet{luck07} \\
Radius (\Rsun) & 5.29$\pm$0.41 & This work \\
   & 5.33$\pm$0.51 & \citet{jones11} \\
Luminosity (\Lsun) & 14.1$\pm$1.8 & This work \\
   & 14.4 & \citet{jones11} \\
   & 12.6 & \citet{mass08} \\
Age (Gyr) & $1.88^{+0.76}_{-0.48}$ & This work \\
\enddata
\end{deluxetable}

\clearpage

%----------------------------------------------------------

\begin{deluxetable}{lllll}
\tabletypesize{\scriptsize}
\tablecolumns{5}
\tablewidth{0pt}
\tablecaption{HD\,33844 Planetary System Parameters }
\tablehead{
\colhead{Parameter} & \multicolumn{2}{c}{Keplerian Fit} & 
\multicolumn{2}{c}{Dynamical Fit} \\
\colhead{} & \colhead{HD\,33844b} & \colhead{HD\,33844c} & 
\colhead{HD\,33844b} & \colhead{HD\,33844c} }

\startdata
\label{planetparams}
Period (days) & 551.4$\pm$7.8 & 916.0$\pm$29.5 & 547.9$\pm$6.4 & 924.3$\pm$32.5 \\
Eccentricity & 0.15$\pm$0.07 & 0.13$\pm$0.10 & 0.16$\pm$0.07 & 0.09$\pm$0.08 \\
$\omega$ (degrees) & 211$\pm$28 & 71$\pm$67 & 190$\pm$62 & 5$\pm$30 \\
$K$ (\ms) & 33.5$\pm$2.0 & 25.4$\pm$2.9 & 32.9$\pm$2.2 & 24.0$\pm$2.2 \\
$T_0$ (BJD-2400000) & 54609$\pm$41 & 54544$\pm$164 & 54578$\pm$50 & 54356$\pm$281 \\
m sin $i$ (\Mjup) & 1.96$\pm$0.12 & 1.75$\pm$0.18 & 1.92$\pm$0.11 & 1.68$\pm$0.16 \\
$a$ (AU) & 1.60$\pm$0.02 & 2.24$\pm$0.05 & 1.59$\pm$0.01 & 2.25$\pm$0.03 \\
\hline
RMS of fit -- AAT (\ms) & 5.9  &   & 9.4  &   \\
RMS of fit -- Keck (\ms) & 7.2  &   & 7.2  &   \\
RMS of fit -- FEROS (\ms) & 10.7  &   & 11.6  &   \\
Total $\chi^2$ (54 d.o.f.) & 65.7 &   &  66.4  &  \\
\enddata
\end{deluxetable}

%  dynamical fit total rms = 7.4
%  keplerian total rms = 7.3

%-------------------------------------------------------------------

\begin{figure}
%\plottwo{hd33844_inner.png}{hd33844_outer.png}
\plottwo{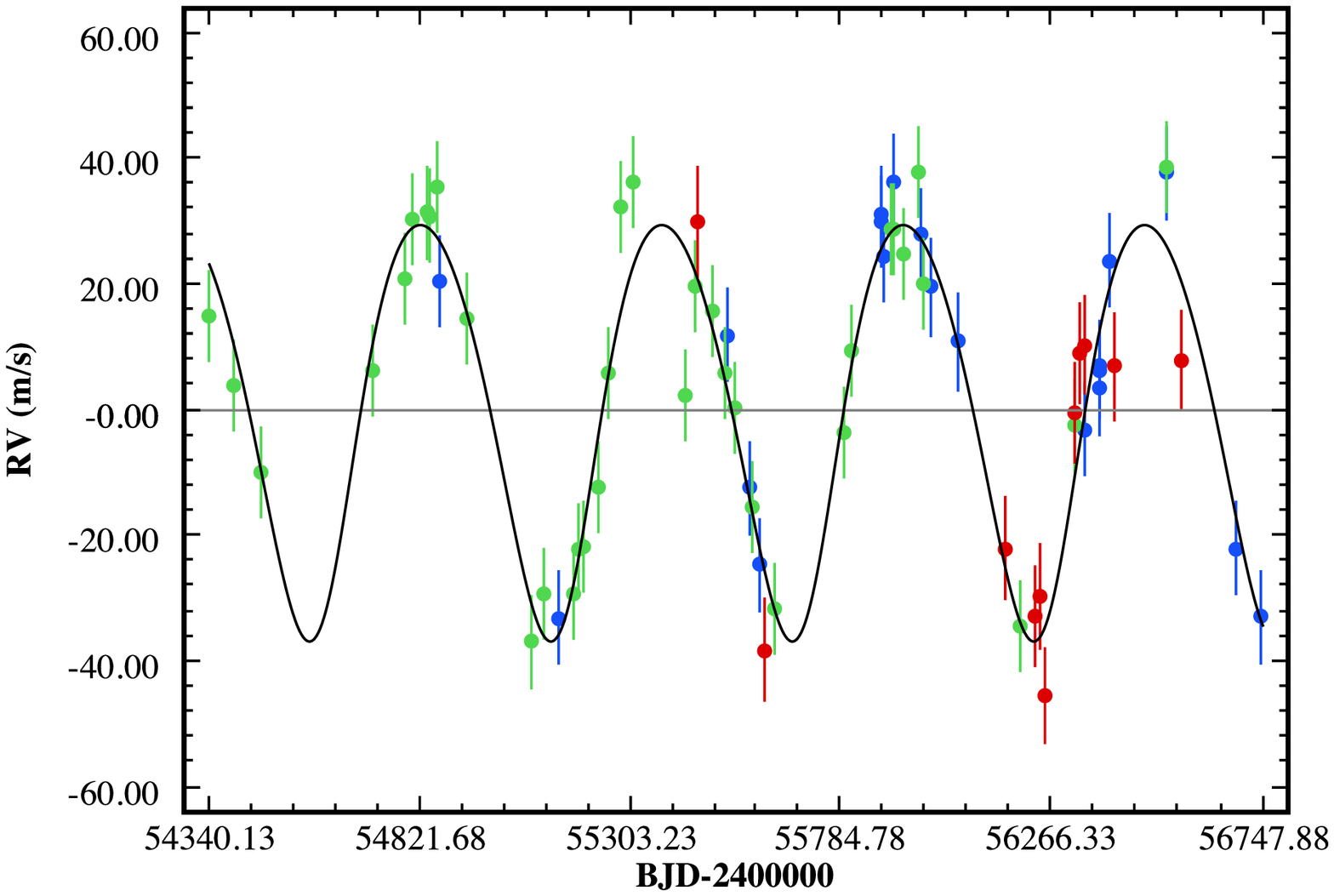}{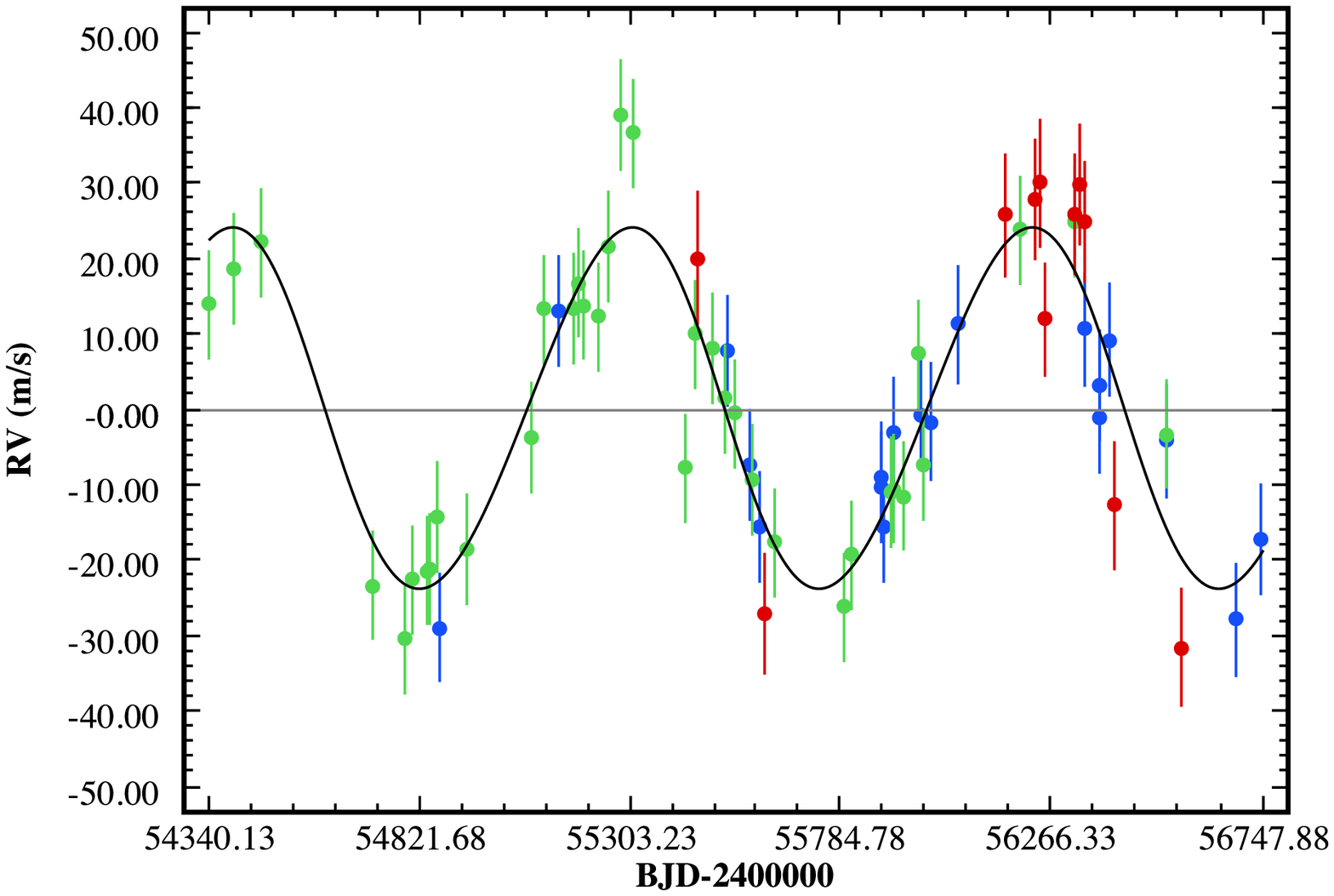}
\caption{Left panel: Data and Keplerian model fit for the inner planet 
HD\,33844b, with the outer planet removed.  Error bars are the 
quadrature sum of the internal uncertainties and 5\,\ms\ of jitter.  
Right panel: Same, but for the outer planet HD\,33844c with the inner 
planet removed.  The total rms about the two-planet Keplerian fit is 
7.3\,\ms.  AAT -- blue, Keck -- green, FEROS -- red. }
\label{individual} 
\end{figure}

%-------------------------------------------------------------------

\begin{figure}
\plotone{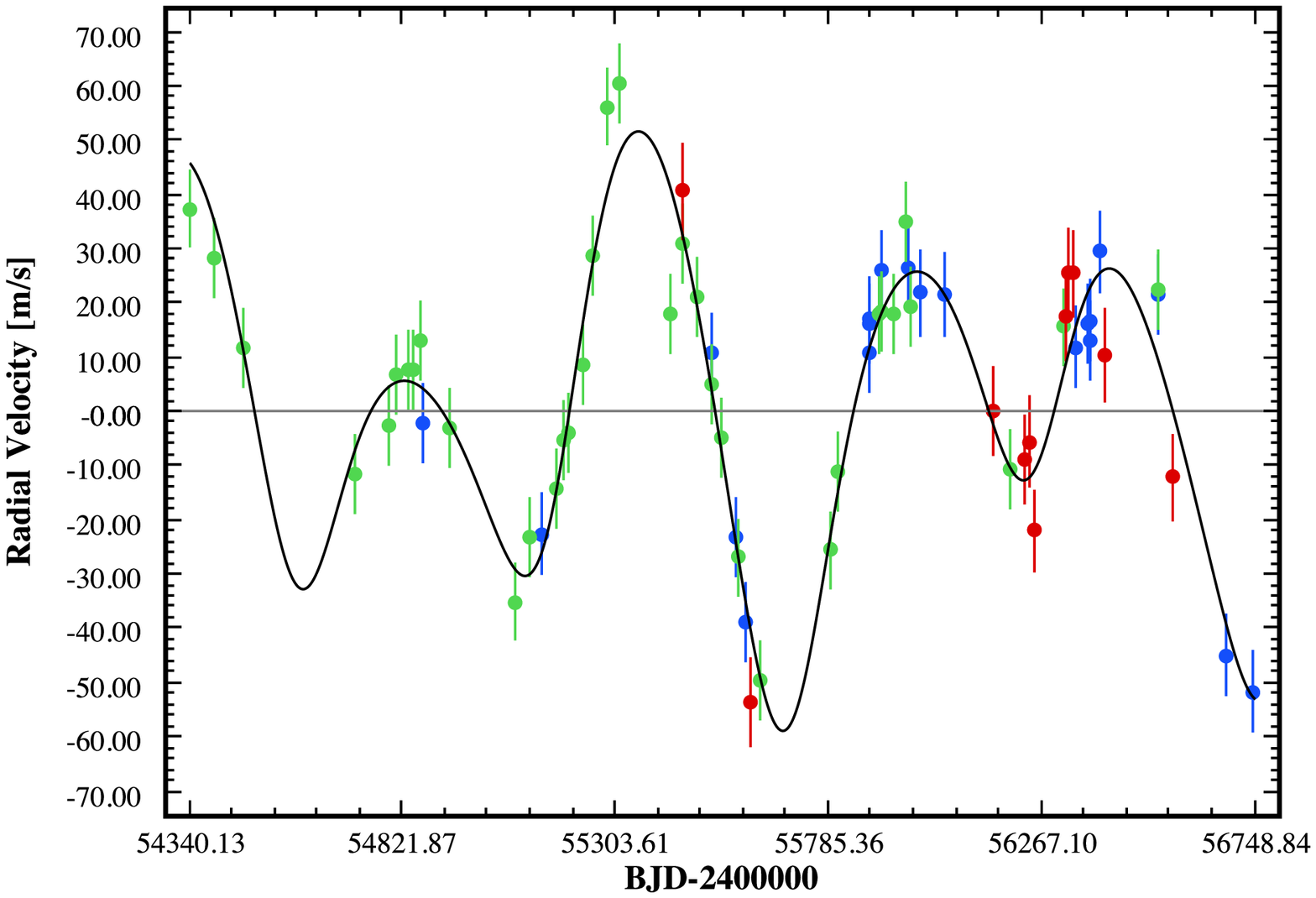}
\caption{Two-planet Keplerian fit for the HD\,33844 system.  The symbols 
have the same meaning as in Figure~\ref{individual}. }
\label{bothplanets}
\end{figure}

%-------------------------------------------------------------------

\begin{figure}
\plotone{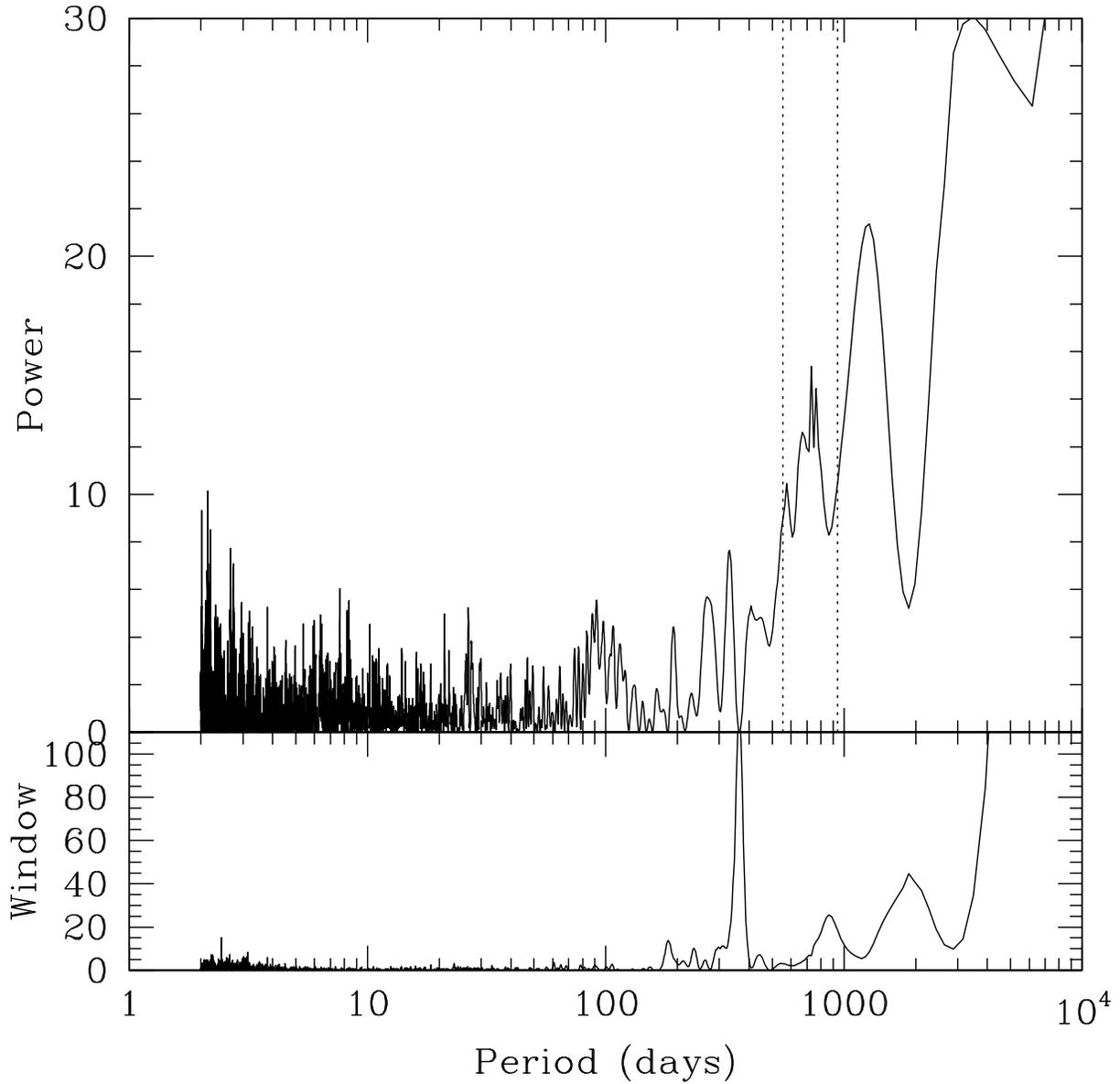}
\caption{Generalised Lomb-Scargle periodogram of ASAS photometry for 
HD\,33844.  A total of 511 epochs spanning 8.8 years reveal no 
periodicities commensurate with the orbital periods of the planets 
(vertical dashed lines). }
\label{asas_pgram}
\end{figure}

%-------------------------------------------------------------------

\begin{figure}
\plottwo{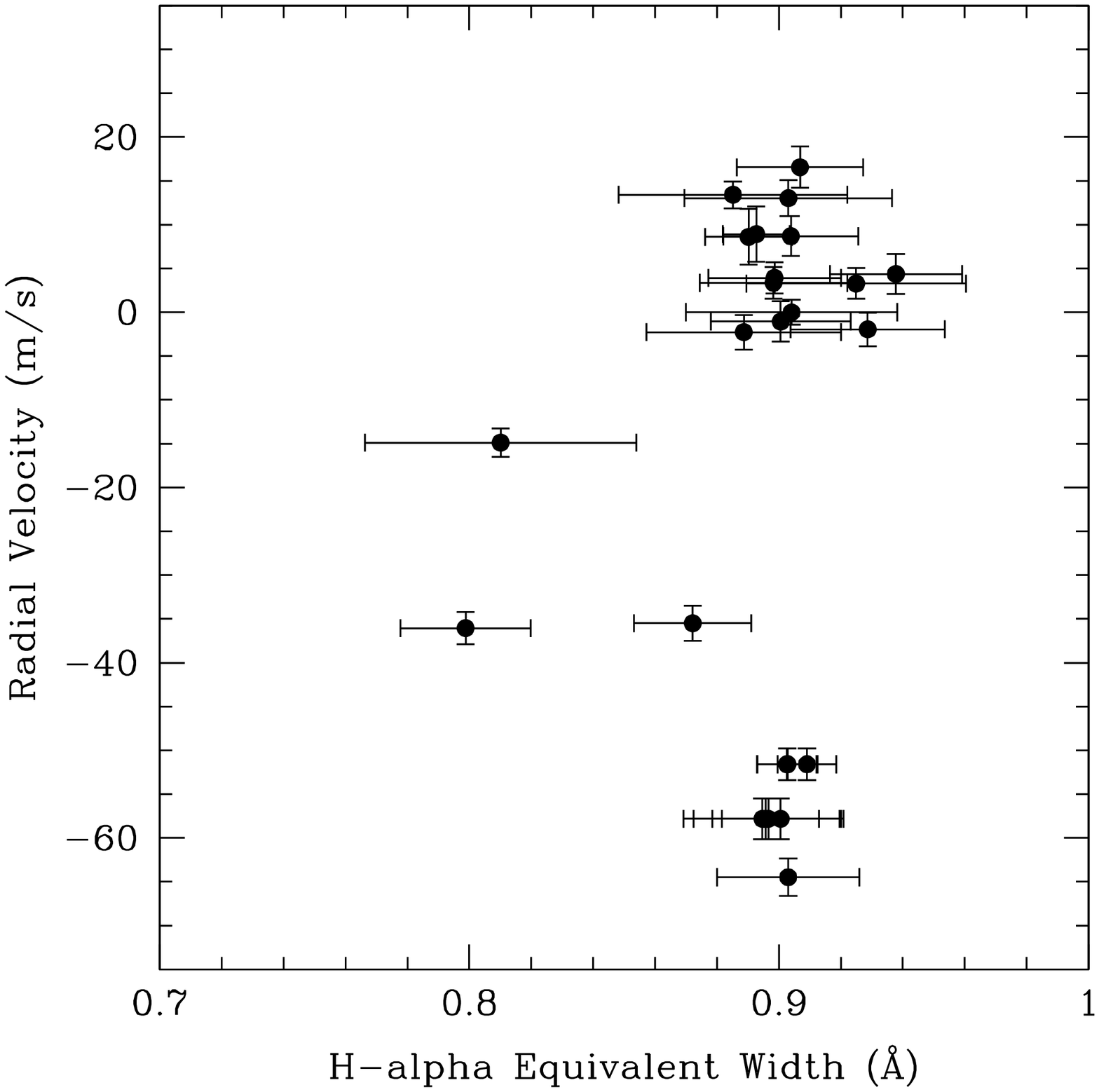}{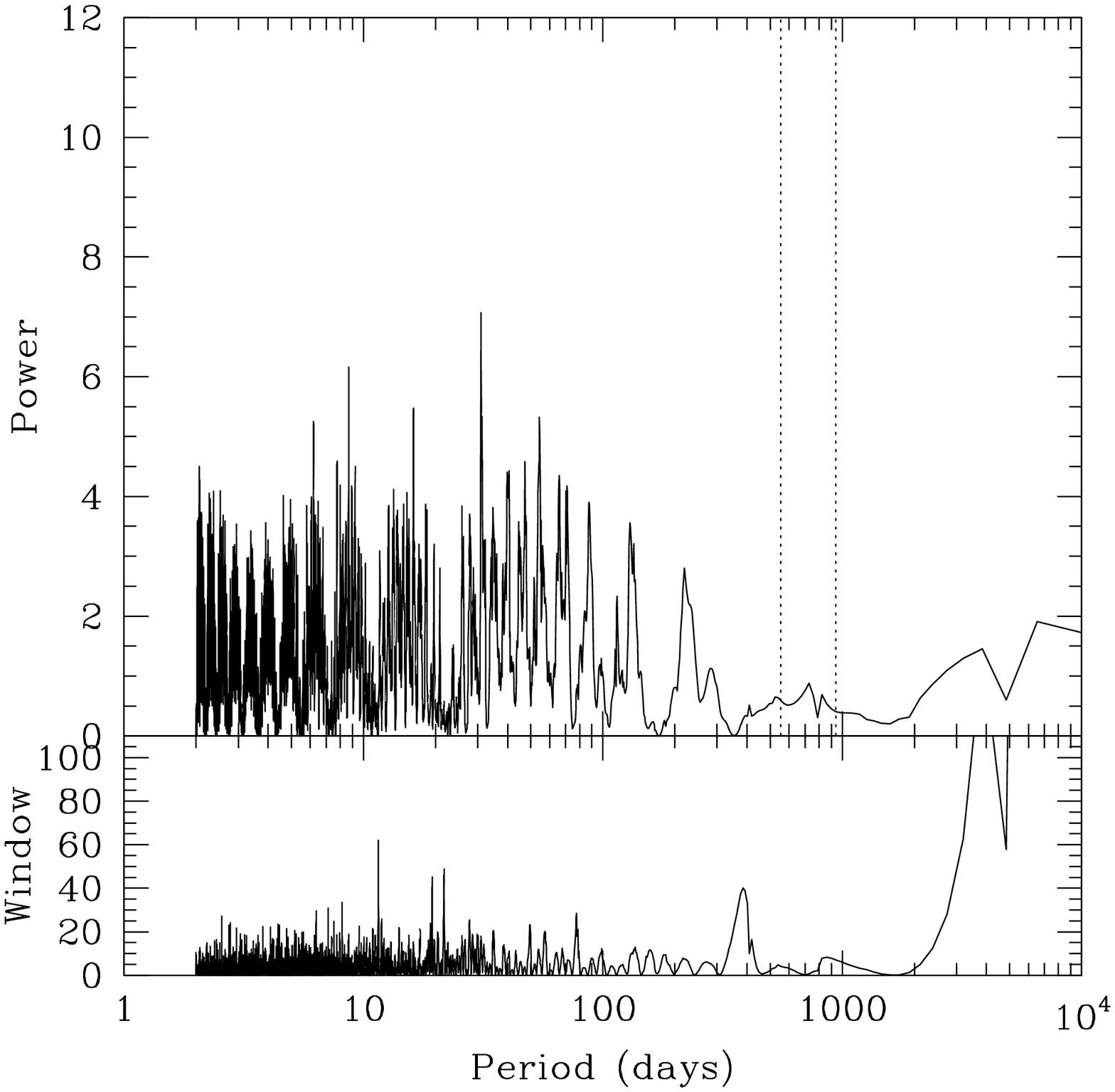}
\caption{Left panel: AAT radial velocities and their H$\alpha$ 
equivalent widths.  No correlations are evident.  Right panel: 
Generalised Lomb-Scargle periodogram of the H$\alpha$ measurements, 
again revealing no significant periodicities.  The orbital periods of 
the planets are marked as vertical dashed lines. }
\label{halpha}
\end{figure}

%-------------------------------------------------------------------

\begin{figure} 
\plotone{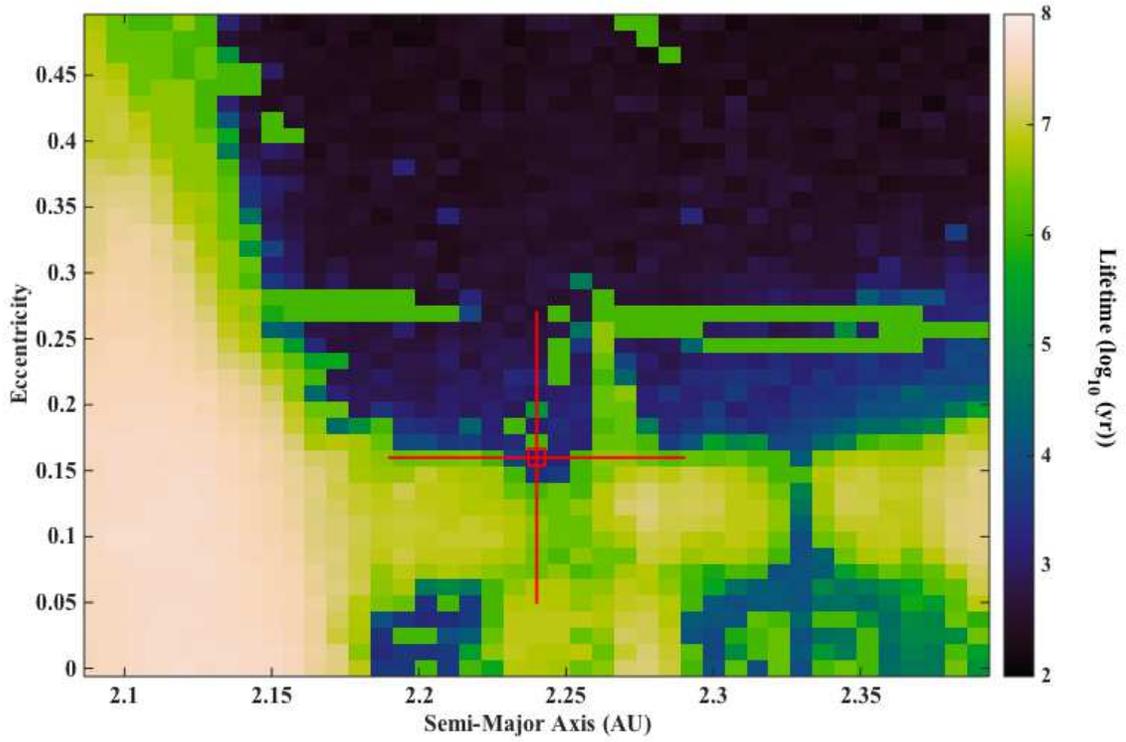} 
\caption{Dynamical stability for the HD\,33844 system as a function of 
the initial semimajor axis and eccentricity of the outer planet.  The 
best-fit orbit for that planet is marked by the open square, and the 
crosshairs show the 1$\sigma$ uncertainties.  Configurations featuring 
eccentricities 1$\sigma$ smaller than the nominal best-fit generally 
remained stable for more than $10^6$ years. }
\label{dynamics1} 
\end{figure}

%-------------------------------------------------------------------

%\begin{figure}
%\plottwo{HD33844Cool_Omega.eps}{HD33844Cool_M_Anom.eps}
%\caption{Left panel: Dynamical stability of the HD\,33844 system as a 
%function of the outer planet's initial argument of periastron $\omega$ 
%and semimajor axis. The best fit and 1$\sigma$ uncertainties are shown 
%by the crosshairs.  While the uncertainties are large, there are 
%significant regions of $>10^6$\,yr stability.  Right panel: Same, but in 
%terms of the outer planet's initial mean anomaly. }
%\label{dynamics2}
%\end{figure}

%-------------------------------------------------------------------

\begin{figure}
\plotone{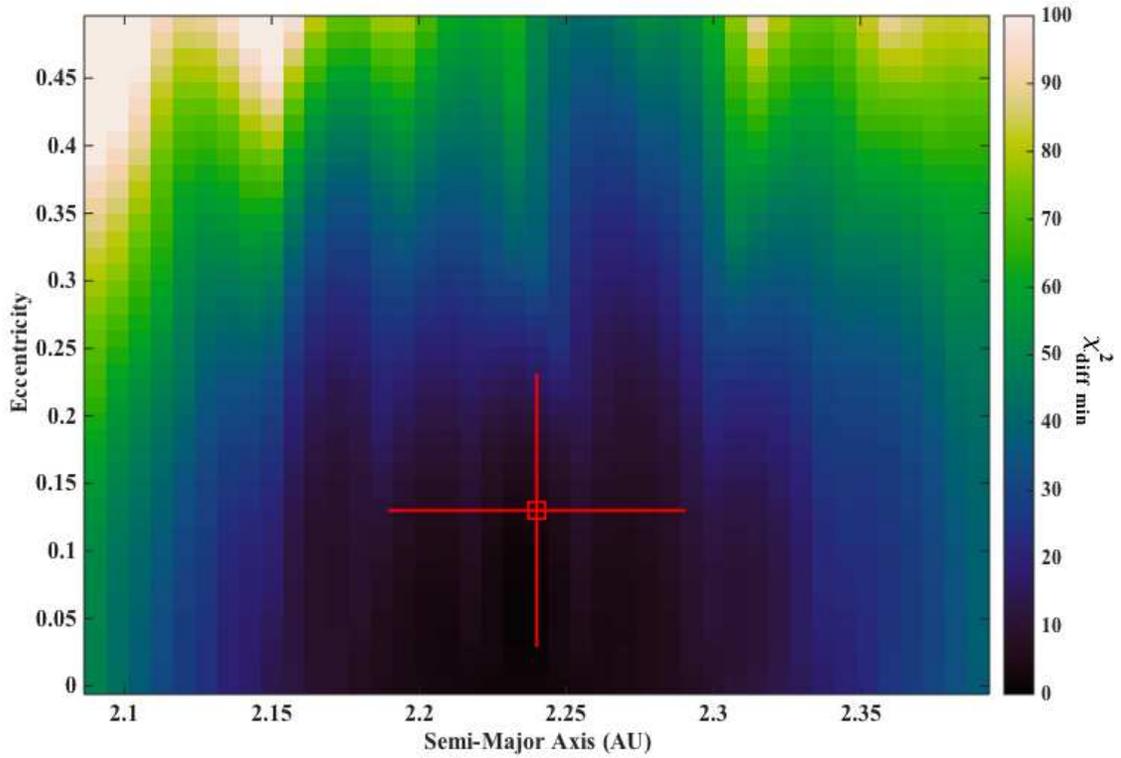}
\caption{$\chi^2$ difference compared to the best fit, for the 126,075 
system configurations tested in Figure~\ref{dynamics1} as a function of 
the initial semi-major axis and eccentricity of HD\,33844c.  At each 
$a-e$ location, we show the minimum value of total $\chi^2$ from the 75 
$\omega$-mean anomaly combinations tested therein (54 degrees of 
freedom). }
\label{chi}
\end{figure}

%-------------------------------------------------------------------

\begin{figure}
\plotone{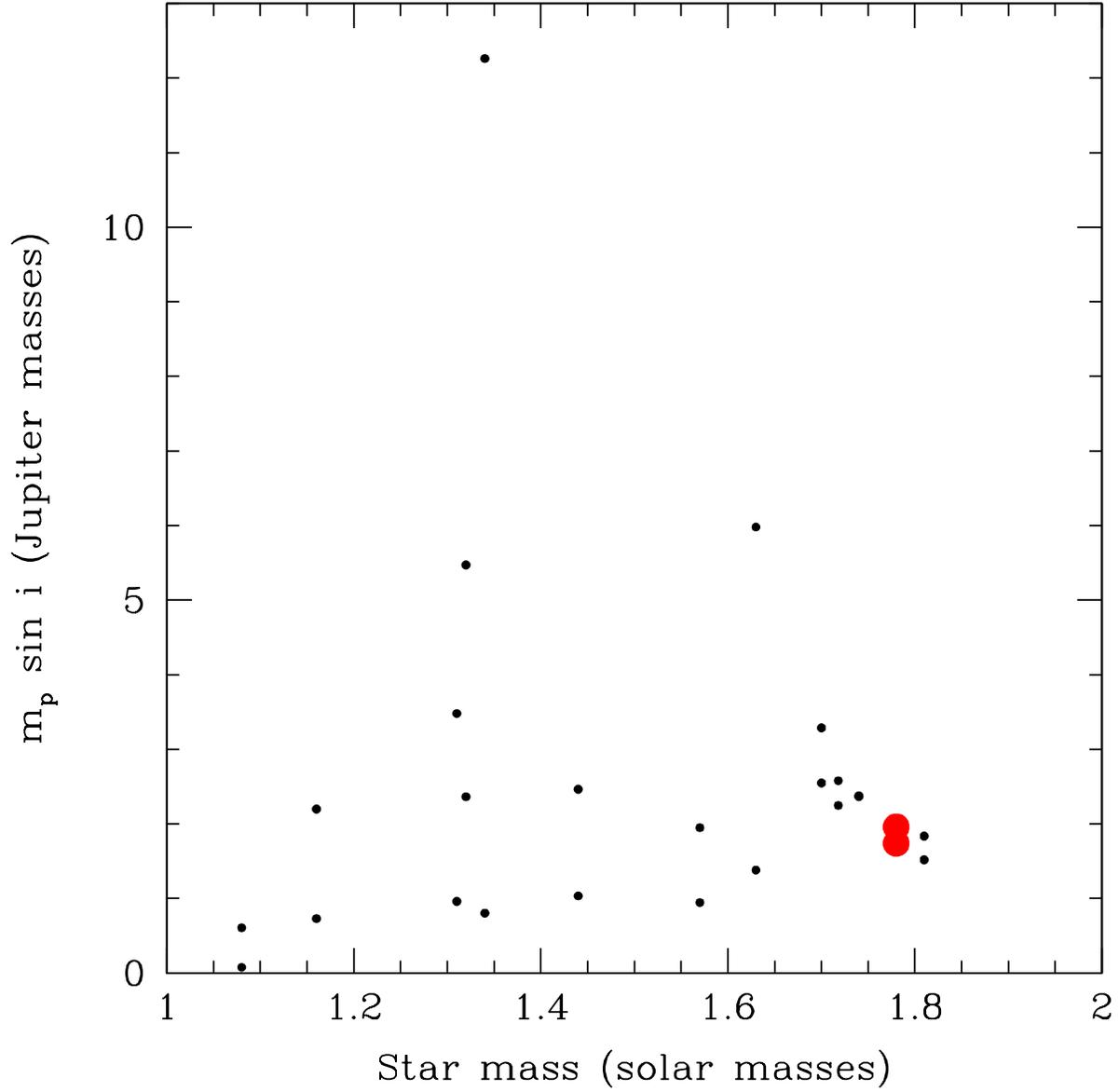}
\caption{Minimum masses (m sin $i$) of planets in multiple systems 
orbiting evolved stars (log $g<4.0$), as a function of stellar mass.  
Only 12 such systems are known; HD\,33844 is shown as large red filled 
circles. }
\label{multis}
\end{figure}

\end{document}